\documentclass[12pt]{article}
\usepackage[colorlinks=true, urlcolor=navyblue, linkcolor=navyblue, citecolor=navyblue]{hyperref}
\usepackage{graphicx,amsmath,amssymb}
\usepackage{indentfirst}
\usepackage[usenames]{color}
\usepackage{epstopdf}
\usepackage{enumerate}
\usepackage{appendix}
\usepackage{lineno}
\usepackage{setspace}
\usepackage{ulem}

\definecolor{navyblue}{rgb}{0,0.08,0.45}
\definecolor{darkred}{rgb}{0.7,0.0,0.0}
\definecolor{darkgreen}{rgb}{0,0.6,0.2}

\newcommand{\beq}{\begin{equation}}
\newcommand{\enq}{\end{equation}}
\newcommand{\beqa}{\begin{eqnarray}}
\newcommand{\beqast}{\begin{eqnarray*}}
\newcommand{\enqa}{\end{eqnarray}}
\newcommand{\enqast}{\end{eqnarray*}}
\newcommand{\nn}{\nonumber}

\newcommand{\bec}{\begin{center}}
\newcommand{\enc}{\end{center}}
\newcommand{\beqo}{\begin{quote}}
\newcommand{\enqo}{\end{quote}}
\newcommand{\bem}{\begin{minipage}}
\newcommand{\enm}{\end{minipage}}

\newcommand{\mbf}{\mathbf}

\newcommand{\req}[1]{(\ref{#1})}

\newcommand{\half}{\textstyle \frac{1}{2}}

\newcommand{\cL}{{\cal L}}

\newcommand{\ga}{\gamma}
\newcommand{\de}{\delta}

\newcommand{\la}{\lambda}
\newcommand{\rh}{\rho}
\newcommand{\si}{\sigma}

\newcommand{\ph}{\phi}

\newcommand{\De}{\Delta}

\newcommand{\La}{\Lambda}

\newcommand{\Si}{\Sigma}

\newcommand{\ud}{\mathrm{d}}

\definecolor{green}{rgb}{0,.5,0}

\begin{document}

\begin{flushright}
{\small SLAC--PUB--16504
}
\end{flushright}

\vspace{50pt}

\begin{center}

{\huge  Universal Effective Hadron Dynamics}

\vspace{6pt}

{\huge   from Superconformal Algebra}
\end{center}

\vspace{30pt}

\centerline{Stanley J. Brodsky}

\vspace{3pt}

\centerline {\it SLAC National Accelerator Laboratory, Stanford
University, Stanford, CA 94309,
USA
}

\vspace{10pt}

\centerline{Guy F. de T\'eramond}

\vspace{3pt}

\centerline {\it Universidad de Costa Rica, 11501 San Pedro de
Montes de Oca, Costa
Rica
}

\vspace{10pt}

\centerline{Hans G\"unter Dosch}

\vspace{3pt}

\centerline{\it Institut f\"ur Theoretische Physik der Universit\"at,}

\centerline{\it   Philosophenweg
16, D-69120 Heidelberg,
Germany
}

\vspace{10pt}

\centerline{C\'edric Lorc\'e}

\vspace{3pt}

\centerline{\it {Centre de Physique Th\'eorique, \'Ecole
polytechnique, CNRS, Universit\'e Paris-Saclay,}}

\centerline{\it {F-91128 Palaiseau, France}
}

\vspace{15pt}

{\small
}

\vspace{60pt}

{\small
\centerline{\href{mailto:sjbth@slac.stanford.edu}{\tt sjbth@slac.stanford.edu}}
}

{\small
\centerline{\href{mailto:gdt@asterix.crnet.cr}{\tt gdt@asterix.crnet.cr}}
}

{\small
\centerline{\href{mailto:dosch@thphys.uni-heidelberg.de}{\tt dosch@thphys.uni-heidelberg.de}}
}

{\small
\centerline{\href{mailto:cedric.lorce@polytechnique.edu}{\tt cedric.lorce@polytechnique.edu}}
}


\newpage

\vspace{15pt}

\begin{abstract}

\vspace{15pt}

An effective supersymmetric  QCD  light-front Hamiltonian for hadrons composed of light quarks, which includes a spin-spin interaction between the hadronic constituents,
is constructed by embedding superconformal quantum mechanics into AdS space. A specific breaking of conformal symmetry inside the graded algebra determines a unique effective quark-confining potential for light hadrons, as well as remarkable connections between the meson and baryon spectra. The results  are consistent with the empirical features of the light-quark hadron spectra, including a universal mass scale for the slopes of the meson and baryon Regge trajectories and a zero-mass pion in the limit of massless quarks.  Our analysis is consistently applied to the excitation spectra of the $\pi, \rho, K, K^*$ and $\phi$   meson families as well as to the $N, \Delta, \Lambda, \Sigma, \Sigma^*, \Xi$ and $\Xi^*$ in the baryon sector. We also predict the existence of tetraquarks which are degenerate in mass with baryons with the same angular momentum.   The mass of light hadrons is expressed in a universal and frame-independent decomposition in the semiclassical approximation described here. 

\end{abstract}

\newpage

\tableofcontents

\section{Introduction \label{introduction}}

Semiclassical approximations for quantum field theory are particularly  useful for hadron physics since the detailed dynamics of QCD is far from being fully understood~\cite{Jaffe:2000ne, Brambilla:2014jmp}. QCD is a  strongly interacting relativistic quantum field theory  where the presence of light quarks and the implementation of color confinement pose special problems~\cite{Gribov:1998kb,Gribov:1999ui,Dokshitzer:2004ie}. The search for semiclassical bound-state equations in gauge theories has received a significant advance through the Maldacena conjecture~\cite{Maldacena:1997re} which states that, under stringent conditions, a strongly interacting relativistic Yang-Mills  conformal quantum field theory is the holographic dual of a classical gravitational theory in a higher-dimensional anti-de Sitter (AdS) space.  This conjecture, the AdS/CFT correspondence, has led to many successful applications in particle, heavy-ion, and solid-state physics~\cite{Ammon:2015wua}.

A very successful phenomenological application of the gauge/gravity correspondence is light-front holographic QCD (LFHQCD)~\cite{Brodsky:2006uqa, deTeramond:2008ht, deTeramond:2013it, Brodsky:2014yha}.  In this case, the holographic variable $z$ in the 5-dimensional classical gravity theory is identified with the boost-invariant transverse separation $\zeta$ between constituents in the light-front quantization scheme \cite{Dirac:1949cp, Brodsky:1997de}. For a system of $N$ constituents, the distance $\zeta$ is that of the ``active'' constituent to the remaining cluster of the $N-1$ constituents.  Such a cluster decomposition is necessary in order to describe baryons in LFHQCD, since there is only a single holographic variable.  This identification does not necessarily imply that the cluster is a tightly bound system; it only requires that essential dynamical features can be described in terms of  the holographic variable $\zeta$. This assumption is supported by the observed similarity between the baryon and meson spectra~\cite{Klempt:2009pi}.

Qualitatively, the observed spectra of both light-quark baryons and mesons show approximately equally spaced parent and daughter trajectories with a common Regge slope~\cite{Klempt:2009pi}. This remarkable structure, especially the equal slopes of the meson and baryon trajectories, suggests the existence of a deeper underlying symmetry.   In the AdS/CFT correspondence~\cite{Maldacena:1997re} the dual quantum field theory is, in fact, a superconformal gauge theory.  Guided by these very general considerations, we will use a simple representation of superconformal algebra to construct semiclassical supersymmetric bound-state equations which are holographically mapped to relativistic  Hamiltonian bound-state equations  in the light front (LF)~\cite{Brodsky:2013ar, deTeramond:2014asa, Dosch:2015nwa}. These wave equations satisfactorily  reproduce  the successful empirical results previously obtained from LFHQCD (see {\it e.g.}~\cite{Brodsky:2014yha}),  with the crucial advantage that additional constant terms in the confinement potential, which are essential for describing the observed phenomenology, are determined from the onset~\cite{Jugeau:2008ds}.

The supersymmetric approach  to hadronic physics also provides unexpected connections  across the heavy-light hadron spectra,  a  sector where one cannot start from a superconformal  algebra because of the strong explicit breaking of conformal symmetry by heavy quark masses~\cite{Dosch:2015bca}.

In our framework, the emerging dynamical supersymmetry between mesons and baryons is not a consequence of supersymmetric QCD at the level of fundamental fields, but the  supersymmetry between the LF bound-state wave functions of mesons and baryons. This symmetry is consistent with an essential feature of color $SU(N_C)$:  a cluster of $N_C-1$ constituents  can be in the same color representation as the anti-constituent; for $SU(3)$ this means $\bf \bar 3 \in  \bf 3 \times \bf 3$ and $\bf  3 \in  \bf \bar3 \times \bf \bar3$~\footnote{This was the basis of earlier attempts~\cite{Miyazawa:1966mfa, Catto:1984wi, Lichtenberg:1999sc} to combine mesons and baryons in supermultiplets.}.

In   AdS$_5$  the positive and negative-chirality projections of the baryon wave functions, the upper and lower spinor components in the chiral representation of Dirac matrices, satisfy uncoupled second-order differential equations with degenerate eigenvalues.  As we shall show in this letter, these component wave functions form, together with the boson wave functions,  the supersymmetric multiplets.

The  semiclassical LF effective theory based on superconformal quantum mechanics also captures   other  essential features of hadron physics that one expects from confined quarks in QCD.  For example,  a massless  pseudoscalar $q \bar q$ bound state --the  pion--  appears in the limit of zero-quark masses,  and a mass scale emerges from a nominal conformal theory.  Moreover, the eigenvalues of the light-front Hamiltonian predict the same slope for Regge trajectories in both $n$,  the radial excitations,  and $L$, the orbital excitations, as approximately observed empirically.  This nontrivial aspect of hadron physics~\cite{Glozman:2007at, Shifman:2007xn}  -- the observed equal slopes of the radial and angular Regge trajectories -- is  also a property of the Veneziano dual amplitude~\cite{Veneziano:1968yb}.

In this letter we will extend our previous analyses and derive a semiclassical  light-front relativistic Hamiltonian based on superconformal algebra and its holographic embedding, which includes a spin-spin interaction between the hadronic constituents.  This extension of our previous results provides a remarkably simple, universal  and consistent description of the  light-hadron spectroscopy and their light-front wavefunctions.   We also predict the existence of tetraquarks which are degenerate with baryons with the same angular momentum. The tetraquarks are bound states of the same confined color-triplet diquark and anti-diquark clusters which account for baryon spectroscopy; they are required to complete the supermultiplet structure  predicted by the superconformal algebra.

\section{Supersymmetric quantum mechanics and hadron physics}

We start this section with a short recapitulation of our previous applications~\cite{deTeramond:2014asa, Dosch:2015nwa} of superconformal quantum mechanics to light front holographic QCD following Ref.~\cite{Fubini:1984hf}. This novel approach to hadron physics not only allows a treatment of nucleons which is completely analogous to that of mesons, but  it also captures  the essential properties  of the  confinement dynamics of light hadrons  and provides a theoretical foundation for the observed similarities between mesons and baryons. The superconformal algebraic approach is then extended to include the spin-spin interactions of the constituents and the contribution to the hadron spectrum from quark masses.

\subsection{Construction of the Hamiltonian from the superconformal algebra}

We briefly review the essential features of the simplest superconformal graded algebra~\cite{Fubini:1984hf}  in one dimension, {\it conf}({$\mathbb R^1$}). It is  based on the  generators of translation, dilatation and the special conformal transformation $H$, $D$ and $K$, respectively. By introducing the supercharges $Q$, $Q^\dagger$, $S$ and $S^\dagger$, one constructs the extended algebraic structure~\cite{Fubini:1984hf, Haag:1974qh}  with the relations
\begin{align} \label{susy-extended}  \nn
\half\{Q,Q^\dagger\} &= H, & \half\{S,S^\dagger\}&=K, \\ 
\{Q,S^\dagger\} &= { f} \, { {\mathbf I}} - B + 2 i D, &
\{Q^\dagger,S\} &= { f} \, {{\mathbf I}} - B - 2 i D,
\end{align}
where $f$ is a real number, ${\mathbf I}$ is the identity operator,
$B=\frac{1}{2}[\psi^\dag,\psi]$ is a bosonic operator with
$\{\psi,\psi^\dag\}={\mathbf I}$, $S=x\,\psi^\dag $ and
$S^\dag=x\, \psi $. The operators $H$, $D$ and $K$ satisfy the
conformal algebra 
\beq [H,D]= iH, \qquad [H,K]= 2 i D, \qquad
[K,D]=-i K . 
\enq

The fermionic operators can be conveniently represented in a spinorial space $S_2={\cal L}_2({\mathbb R^1})\otimes {\mathbb C^2}$ as 2$\times$2 matrices 
\beq
Q = \left(\begin{array}{cc}
0&q\\
0&0\\
\end{array}
\right) ,
\quad Q^\dagger=\left(\begin{array}{cc}
0&0\\
q^\dagger&0
\end{array}
\right),
\enq
\beq
S = \left(\begin{array}{cc} 
0& x\\
0&0
\end{array}
\right), \quad  S^\dagger=\left(\begin{array}{cc}
0&0\\
x&0\\
\end{array}
\right) , \enq 
    with 
      \beq \label{q} q =-\frac{\ud}{\ud x} + \frac{f}{x},
\qquad
 q^\dagger = \frac{\ud}{\ud x}  + \frac{f}{x}.
\enq

Following the analysis of Fubini and Rabinovici~\cite{Fubini:1984hf}, which extends the treatment of the conformal group by de Alfaro, Fubini and Furlan~\cite{deAlfaro:1976je} to supersymmetry, we construct ~\cite{deTeramond:2014asa, Dosch:2015nwa} a generalized Hamiltonian from the supercharges
\beq \label{R}
R_\la =Q+\la\,S =\left(\begin{array}{cc}
0&-\frac{\ud}{\ud x} + \frac{f}{x}+\la\,x\\
0&0\\
\end{array}
\right) ,
\enq
\beq  \label{Rdag}
\quad R_\la^ \dagger=Q^\dagger + \la S^\dagger =\left(\begin{array}{cc}
0&0\\
\frac{\ud}{\ud x} + \frac{f}{x}+\la\,x&0
\end{array}
\right),
\enq
namely
\beq \label{G}
G = \{R_\la, R_\la^\dagger\}.
\enq
Since the dimensions of the generators $Q$ and $S$ are different, a scale $\la$, with dimensions of mass squared,  is introduced in the Hamiltonian in analogy with the earlier treatment of conformal quantum mechanics given in Ref.~\cite{deAlfaro:1976je}.   As shown by de Alfaro, Fubini and Furlan~\cite{deAlfaro:1976je}, the conformal symmetry of the action is retained despite the presence of a mass scale in the Hamiltonian.

The supercharges and the new Hamiltonian $G$  satisfy, by construction,  the relations:
\beq  
\label{newrel} \{R_\la^\dagger,R_\la^\dagger\}
=\{R_\la,R_\la\}=0,\qquad[R_\la,G]= [R_\la^\dagger,G]=0,
\enq
which, together with Eq. \req{G}, close a graded algebra $sl(1/1)$, as in Witten's supersymmetric quantum mechanics~\cite{Witten:1981nf}.  Since the Hamiltonian $G$ commutes with $R^\dagger_\la$, it  follows that the states  $\vert \phi \rangle$ and $R^\dagger  \vert \phi \rangle$ have identical eigenvalues. Furthermore, it follows that if $|\ph_0 \rangle$ is an eigenvalue of $G$ with zero eigenvalue, it is annihilated by the operator $R_\la^\dagger$:
\beq  \label{anni} 
R_\la^\dagger|\ph_0 \rangle = 0.
\enq

In matrix representation Eq. \req{G} is given by
\beq
G = 2 H + 2 \la^2 K + 2 \la \left( f\, \mbf{I} - \si_3 \right),
 \quad \mbox{with} \quad \si_3 = \left(\begin{array}{cc} 1&0 \\ 0 & -1 \end{array}\right).
\enq
The new Hamiltonian is diagonal, with elements:
\beqa \label{Gsc}
G_{11}&=& - \frac{\ud^2}{\ud x^2} + \frac{4 (f + \half)^2 - 1}{4
 x^2} + \la^2 \,x^2+ 2 \la\,(f-\half),\\
\label{Gsd} G_{22}&=& - \frac{\ud^2}{\ud  x^2} + \frac{4 (f-\half)^2 -
1}{4  x^2}+  \la^2 \, x^2+ 2
 \la\,(f+\half).
\enqa

These equations have the  same  structure as  the second order wave equations in AdS space, which follow from a linear Dirac equation with a multiplet structure composed of positive and negative-chirality components~\cite{deTeramond:2013it, deTeramond:2014asa}.  Mapping to light-front physics, one identifies the conformal variable $x$ with $\zeta$, the boost-invariant LF separation of the constituents~\footnote{In general, the invariant light-front variable of the $N$-quark bound state is $\zeta = \sqrt{\frac{x}{1-x}}\, \Big\vert \sum_{j=1}^{N-1} x_j \mbf{b}_{\perp j} \Big\vert$,  where $x$ is the longitudinal momentum fraction of the active quark, $x_j$ with $j =1, 2, \cdots, N-1$,  the momentum fractions associated with the $N-1$ quarks in the cluster, and $ \mbf{b}_{\perp j}$ is the transverse positions of the spectator quarks in the cluster relative to the active one~\cite{Brodsky:2006uqa}. For a two-constituent bound-state $\zeta = \sqrt{x(1-x)}\, \vert\mbf{b}_{\perp}\vert$, which is conjugate to the invariant mass $\frac{\mbf{k}_\perp^2}{x(1-x)}$.}$^,$\footnote{These equations  are analogous to the LFHQCD relation of Eqs.~(5.28) and~(5.29) to Eqs.~(5.32) and~(5.33) in Ref.~\cite{Brodsky:2014yha}.   In the case of fermions, the maximal symmetry of AdS was broken in LFHQCD by the introduction of an {\it ad hoc}  Yukawa-like term in the AdS action~\cite{Abidin:2009hr}. This is unnecessary using superconformal algebra.}.

The operator $G_{22}$ agrees with the LF Hamiltonian of the positive-chirality projection; similarly,  the operator  $G_{11}$ acts on the negative-chirality component. The positive-chirality component  $\psi_+(\zeta) \sim \zeta^{\frac{1}{2} + L} e^{-\la \zeta^2/2} L_n^L(\la \zeta^2)$ has orbital angular momentum  $L_B= f-\half$  and it is the leading twist solution; the  negative-chirality component $\psi_-(\zeta) \sim  \zeta^{\frac{3}{2} + L} e^{-\la \zeta^2/2}  L_n^{L+1}(\la \zeta^2)$ has $L_B+1$. The total nucleon wave function is the plane-wave superposition~\cite{deTeramond:2013it, deTeramond:2014asa}~\footnote{In AdS$_5$ the Dirac matrix corresponding to the fifth (the holographic) variable is proportional to the matrix $\ga_5$,  $\Gamma_z = i \gamma_5$.   Eq.~\req{Psi} is, in general,  the solution of the linear Rarita-Schwinger equation in AdS$_5$ space which determines the relative normalization of both components~\cite{deTeramond:2013it, deTeramond:2014asa}.}
 \beq \label{Psi}
 \Psi(x^\mu, \zeta) =  e^{i P \cdot x} \left[ \psi^+(\zeta) \tfrac{1}{2}  \left(1 + \gamma_5) u({P}\right) +    \psi^-(\zeta) \tfrac{1}{2}  \left(1 - \gamma_5\right) u({P})\right],
 \enq 
where $u(P)$ is a Dirac spinor of a free nucleon with momentum $P$ in four-dimensional Minkowski space~\cite{deTeramond:2013it, Brodsky:2014yha}. Both components have identical normalization~\cite{deTeramond:2014asa}, and thus the nucleon spin is carried by the LF orbital angular momentum~\cite{Brodsky:2014yha}~\footnote{The equality of the normalization of the $L=0$ and $L=1$  components   is also predicted by the Skyrme model~\cite{Brodsky:1988ip}.}.

The operator $G_{11}$ is also the LF Hamiltonian of a meson with angular momentum $J=L_M=f+\half $~\footnote{Compare Eq.~(5.2) with Eq.~(5.5) in Ref.~\cite{Brodsky:2014yha}.}. The eigenfunctions of $G_{11}$ and $G_{22}$ are related by the fermionic operators $R_\la$ and $R^\dagger_\la$,  Eqs. \req{R} and \req{Rdag}; these supercharges can be interpreted as operators which transform baryon into meson  wave functions and vice-versa~\cite{Dosch:2015nwa}. The operator $G_{11}$ is thus  the Hamiltonian for mesons, and $G_{22}$  is the Hamiltonian for the positive-chirality component, the leading-twist baryon wave function.

The eigenfunctions of the Hamiltonian $G_{11}$ are  $\phi_{n,L}(z) \sim \zeta^{1/2 +L} e^{- \la  z^2/2} L_n^L(\la z^2)$, with $L= f+\half$.   Using the relations
$n L_n^\nu(x) = (n + \nu)  L_{n-1}^\nu(x) - x  L_{n-1}^{\nu+1}(x)$ and $  L_n^{\nu-1}(x) =  L_n^\nu(x) -  L_{n-1}^\nu(x)$  between the associated Laguerre polynomials we find 
\beq \label{Rdagphi} 
R_\la ^\dagger \vert \phi^M_{n,L} \rangle = 2 \sqrt \la (n+ L)^{1/2}  \vert \phi^B_{n,L-1} \rangle,
\enq
where 
\beq 
|\phi^M_{n,L} \rangle=\left(\begin{array}{c} \phi_{n,L}\\ 0 \end{array}\right), \quad \quad
|\phi^B_{n,L-1} \rangle=\left(\begin{array}{c}0\\  \phi_{n,L-1}\end{array}\right). 
\enq 
This shows explicitly  the remarkable relation $L_M = L_B + 1$ which identifies the  orbital angular momenta of the mesons  with their baryon superpartners with identical mass~\cite{Dosch:2015nwa}.

The relation $L_B=f-\half$ shows that $f$ must be positive for baryons, in accordance with the requirement that the superconformal potential $\frac{f}{x}$ in \req{q} be bounded from below~\cite{Fubini:1984hf, deAlfaro:1976je}.  However, for mesons,  the negative value $f=-\half$ leads to angular momentum $L_M=0$, which is allowed  and is consistent with the Hamiltonian $G_{11}$ for mesons~\footnote{In LFHQCD the lowest possible value $L_M=0$ corresponds to the lowest possible value for the AdS$_5$ mass allowed by the Breitenlohner-Freedman stability bound~\cite{Breitenlohner:1982jf}.}. We can therefore regard  Eq.~\req{Gsc} as an extension of the supersymmetric theory with $f>0$ to the negative value $f=-\half$ for mesons. It is clear from Eq. \req{Rdagphi} that the fermion operator $R^\dagger$ annihilates the lowest state corresponding to $n = L = 0$, $R^\dagger  \vert \phi_{n = 0, L = 0} \rangle = 0$,  in accordance with  Eq.~\req{anni}. Thus the pion has a special role in the superconformal  approach to hadronic physics as a unique state of zero mass~\cite{Dosch:2015nwa}.  It also follows from Eq.~\req{Rdagphi} that meson states with $n>0$ and $L=0$, also corresponding to the marginal value  $f=-\half$, are not annihilated by $R^\dagger$. These  states, however, are connected to  unphysical fermion  states with $L=-1$.  These spurious states are eigenstates of the Hamiltonian $G$, Eq. \req{G}, with positive eigenvalues  and their presence seems unavoidable in the supersymmetric construction, since each state with eigenvalue different from zero should have a partner, as dictated by the index theorem~\cite{Witten:1981nf}.

The situation is completely analogous to the case where  conformal symmetry is explicitly and strongly  broken by heavy quark masses.  In this case the superpotential  is no longer  constrained by conformal symmetry and it is basically unknown, but the meson-baryon supersymmetry still holds~\cite{Dosch:2015bca}.  
In particular,  the   $L = 0$ meson states have no supersymmetric baryon partner since they would correspond to  unphysical $L = -1$ states.

\subsection{Holographic embedding and the spin-spin interaction}

The pion and nucleon trajectories can be consistently described by the superconformal algebraic structure mapped to the light front~\cite{Dosch:2015nwa}.  A fundamental prediction is a  massless pion in the limit of zero quark masses. However, there remains a lingering question for the $\rh$ and $\Delta$ trajectories:  in the case of LFHQCD,  it was found necessary  to introduce the concept of half-integer twist~\cite{Brodsky:2014yha, Dosch:2015nwa}  in order to describe the $\Delta$ trajectory~\footnote{The analogous problem does not arise for the meson trajectories since the resulting bound-state equation, from its holographic embedding, depends on the total and orbital angular momentum separately, see  Eq.~(5.2) with~(5.5) in Ref.~\cite{Brodsky:2014yha}.}. For states with $J= L_M+s$, where $s$ is the total quark spin, the spin interaction follows from the holographic embedding of the bound-state equations~\cite{Brodsky:2014yha}; it is not determined by the superconformal construction. This amounts to the modification of the meson Hamiltonian $G_{11} \to G_{11} + 2 \la\,  s$.   In  order to preserve supersymmetry,  one must add the same term to the baryon Hamiltonian $G_{22}$. The  resulting supersymmetric Hamiltonian for mesons and baryons in the chiral limit  is therefore
\beq \label{GS}
G_S=  \{R_\la,R^\dagger_\la\}  +  2 \la \,  s \, \mathbf{I},
\enq
For mesons  $s$ is the total internal quark spin of a meson. The identification of baryons  as bound states of a quark  and a  diquark cluster provides a  satisfactory interpretation of the supersymmetric implementation: in  this  case we can identify $s$ with the spin of the diquark cluster. The spin of the diquark cluster of the $\Delta$ trajectory and the nucleon family with total quark spin $\frac{3}{2}$ must be  $s=1$: it is the natural superpartner of the $\rho$ trajectory. For the nucleon family with total quark spin $\half$, the cluster  is, in general, a superposition  of spin  $s=0$ and $s=1$. Since the nucleon trajectory is the natural partner of the $\pi$ trajectory, we have to choose the cluster spin $s=0$ to maintain supersymmetry. In general, we  take $s$ as the smallest possible value compatible with the quantum numbers of the  hadrons and  the Pauli principle. This procedure reproduces the agreement with the empirical baryon spectrum obtained in our previous treatments without  the unsatisfactory feature of introducing half-integer twist;  all twists and orbital angular momenta are integers.

In the case of mesons, the lowest mass state of the vector meson family, the $I=1$ $\rho$ (or the $I=0$ $\omega$ meson) is  annihilated by the fermion operator $R^\dagger$, and it has no baryon superpartner.  This is possible, even though the $\rho$ is a massive particle in the limit of zero quark masses, since the effect of the spin term $2 \la s$ in the new Hamiltonian  is an overall shift of the mass scale without a modification of the LF wavefunction. The action of the fermion operator is thus the same as for the pseudoscalar meson family.

To summarize: The meson wave function $ \phi_M(L_M)$, with LF orbital angular momentum $L_M$ and quark spin $s$, and the positive-chirality (leading-twist) component wave function $\psi_{B+}$  of a baryon, with cluster spin $s$ and orbital angular momentum $L_B=L_M-1$, are part of the supermultiplet
\beq \label{MBplet}
\vert \phi_H\rangle = \left( \begin{array}{c} \phi_M(L_M) \\ \psi_{B+}(L_B=L_M-1) \end{array} \right),
\enq
with equal mass.  The supercharge $R_\lambda^\dagger$ acts on the multiplet \req{MBplet} and transforms the meson wave function into the corresponding baryon wave function. The meson and baryon mass spectra resulting from the Hamiltonian \req{GS} are given by the simple formul\ae:
\beqa \label{hadchiral}  
\mbox{Mesons } && M^2 = 4 \la (n+L_M)+ 2 \la \,  s ,\\
\label{mesfin}\mbox{Baryons} && M^2=4 \la(n+L_B+1) + 2 \la \,  s .
\enqa
As discussed below Eq. \req{GS},  $s$ is the internal quark spin for mesons  and the lowest possible value of the cluster spin for baryons  .

We are working in a semiclassical approximation;  therefore hadron states are described by wave functions in the Hilbert space $\cL_2(R_1)$,  where the  variable is the boost-invariant light-front transverse separation $\zeta$. The generators of the symmetries are operators in that Hilbert space. Since the wave functions (and spectra) are equal for hadrons and anti-hadrons, the superpartner of the meson is a baryon as well as the corresponding antibaryon. 

In order to interpret these results for hadron physics,  we shall assume that the constituents of  mesons and baryons are quarks or antiquarks with the well-known quantum numbers. Thus the fermion operator $R_\lambda^\dagger$ is interpreted as the transformation operator of a  single constituent (quark or antiquark)  into an anti-constituent cluster  in the conjugate color representation.

\subsection{Tetraquarks}

\begin{figure}
\setlength{\unitlength}{1mm}
\begin{center}
\begin{picture}(60,30)(15,0)
\put(20,30){\circle{20}} \put(20,33){\circle*{2}}
\put(20,27){\circle{2}}
 \put(50,30){\circle{20}}
\put(48,33){\circle{2}} \put(52,33){\circle{2}}
\put(50,27){\circle{2}}
 \put(14,19){$\phi_M,\;L_B+1$}
\put(43,19){$\psi_{B+},\;L_B$}
 \put(22,33){\vector(1,0) {24}}
\put(32,35){$R_\lambda^\dagger$}

 \put(50,7){\circle{20}}
\put(48,10){\circle{2}} \put(52,10){\circle{2}}
\put(50,4){\circle{2}} \put(43,-4){$\psi_{B-}, \;L_B+1$}
\put(80,7){\circle{20}} \put(78,10){\circle{2}}
\put(82,10){\circle{2}} \put(78,4){\circle*{2}}
\put(82,4){\circle*{2}} \put(73,-4){$\phi_T,\;L_B$}
\put(52,4){\vector(1,0) {24}} \put(62,6){$R_\lambda^\dagger$}
\end{picture}
\end{center}
\caption{\label{tetra} \small The supersymmetric quadruplet $\{\phi_M, \psi_{B+}, \psi_{B-},\phi_T\}$. Open circles represent quarks, full circles antiquarks. The tetraquark has the same mass as its baryon partner in the multiplet. Notice that the LF angular momentum of the negative-chirality component wave function of a baryon $\psi_{B-}$ is one unit higher than that of the positive-chirality (leading-twist) component $\psi_{B+}$.}
\end{figure}
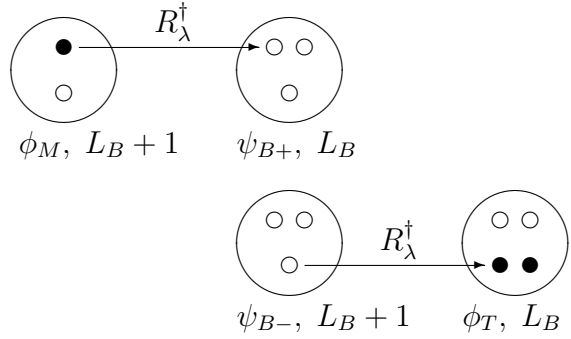

The supersymmetric states introduced in the previous section do not form a complete supermultiplet, since the negative-chirality component wave function of the baryon has not yet been assigned to its supersymmetric partner.   We can complete the supersymmetric multiplet by applying the fermion operator $R_\la^\dagger$ to the negative-chirality component baryon wave function and thus obtain a bosonic state.   As noted above, the operator $R_\la^\dagger$ can be interpreted as transforming a constituent into a
two-anti-constituent cluster in the same color representation as the constituent. It transforms a quark into an anti-diquark in color representation $3$ and an antiquark into a diquark in color representation $\bar 3$.   Therefore  the operator $R_\la^\dagger$ applied to the negative-chirality component of a baryon will give a tetraquark wave function,  $\phi_T= R_\la^\dagger\, \psi_{B-}$, a bound state of a diquark and an anti-diquark cluster  as depicted in Fig.~\ref{tetra}.

The negative-chirality component of a baryon, $\psi_{B-}$, has LF angular momentum $L_B+1$ if its positive-chirality component partner has LF angular momentum $L_B$.  Since $R_\la^\dagger$ lowers the angular momentum by one unit, the angular momentum of the corresponding tetraquark is  equal to that of the positive-chirality component of the baryon, $L_T=L_B$. The full supersymmetric quadruplet representation thus consists of two fermion wave functions, namely the positive and negative-chirality components of the baryon spinor wave function $\psi_{B+}$ and $\psi_{B-}$, plus two bosonic wave functions, namely the meson $\phi_B$ and the tetraquark $\phi_T$.  These states can be arranged as a $2 \times2$ matrix: \beq
\left(\begin{array}{cc}
\phi_M{(L_M = L_B+1)} &\psi_{B-}{(L_B+ 1)}\\
\psi_{B+}{(L_B)} &\phi_T{(L_T = L_B)}
\end{array}
\right), 
\enq
 on which the symmetry generators~\req{susy-extended} and the Hamiltonian~\req{GS} operate~\footnote{It is interesting to note that in Ref.~\cite{Catto:1984wi} mesons, baryons and tetraquarks  are also hadronic states within  the same multiplet.}.

According to \req{GS} the total quark spin of all states must be the same. Furthermore we have to take into account that the diquark as a two-fermion state has to be totally antisymmetric. The colour indices are antisymmetric and therefore the spin and isospin of a cluster of two light quarks ($u,\,d$) are correlated. Quark spin $s=0$ goes together with isospin $I=0$,  and $s=1$ entails  $I=1$. The resulting cluster configurations for several families of baryons and their tetraquark partners are displayed in Table \ref{tetratable}.

{\renewcommand{\arraystretch}{1.1}  
\begin{table}[h]
\begin{center}
\begin{tabular}{| l | ccc|ccc|}
\hline
& \multicolumn{3}{c|}{Baryon} & \multicolumn{3}{c|}{Tetraquark}\\
& & $s$ & $I$ & &$s$ &$I$ \\
\hline
N-& $q$ & $\half$ & $\half$ & $(\bar q \bar q)$ & 0 & 0 \\
fam. & $(qq)$ & 0 & 0 & $(qq)$ & 0 & 0 \\
\hline
$\De$-& $q$ & $\half$ & $\half$ & $(\bar q \bar q)$ & 0 & 1 \\
fam. & $(qq)$ & 1 & 1 & $(qq)$ & 1 & 1 \\
\hline
$\La$-& $s$ & $\half$ & $0$ & $(\bar s \bar q)$ & 0 & $\half$ \\
fam. & $(qq)$ & 0 & 0 & $(qq)$ & 0 & 0 \\
\hline
$\Si$-& $q$ & $\half$ & $\half$ & $(\bar q \bar q)$ & 0 & 0\\
fam. & $(sq)$ & 0 & $\half$ & $(sq)$ & 0 & $\half$ \\
\hline
$\Si^*$-& $s$ & $\half$ & $0$ & $(\bar s \bar q)$ & 0 & $\half$ \\
fam. & $(qq)$ & 1 & 1 & $(qq)$ & 1 & 1 \\
\hline
$\Xi$-& $s$ & $\half$ & $0$ & $(\bar s \bar q)$ & 0 & $\half$ \\
fam. & $(sq)$ & 0 & $\half$ & $(sq)$ & 0 & $\half$ \\
\hline
$\Xi^*$-& $s$ & $\half$ & $0$ & $(\bar s \bar q)$ & 0 & $\half$ \\
fam. & $(sq)$ & 1 & $\half$ & $(sq)$ & 1 & $\half$ \\
\hline
\end{tabular}
\end{center}
\caption{\small \label{tetratable} Quantum numbers of the constituents and constituent clusters of different baryon families and their supersymmetric tetraquark partners}
\end{table}

The quantum numbers of the tetraquark itself can be easily calculated from the ones of the two constituent clusters.  Since the relative angular momentum of the two diquarks in the tetraquark is equal to the angular momentum $L_B$ of the positive-chirality component of the baryon, and since the tetraquark consists of an even number of antiquarks, its parity is $(-1)^{L_B}$.

The leading-twist component of the nucleon has $L_B=0, s=0$. Thus its tetraquark partner consists of a diquark and anti-diquark, both with $s=0$; therefore its isospin is $ I=0$. The parity must be $P=+$,  since it has $L=0$ and it consists of two particles and two antiparticles.  A candidate for such a state is the $f_0(980)$. For the partner of the $\De$ resonance we must have $s=1$, it therefore consists of a diquark with  $I=1,\; s=1$ and an anti-diquark with $I=0,\;s=0$. The resulting quantum numbers are $I=1, s=1$ and $P=+$;  the $a_1(1260)$ is a candidate.   The first $L$ excitation of the nucleon is the $N^{3/2-}(1520)$ and $N^{1/2-}(1535)$ pair.  Its tetraquark partner consists of two $I=0,  s=0$ clusters, and thus its quantum numbers are $I=0, J^P =0,1^-$;  candidates are the $\omega(1420)$ and $\omega(1650)$ --or the mixing of these two states.

\subsection{Inclusion of quark masses and comparison with experiment \label{qm}}

We have argued in~\cite{Brodsky:2014yha} that the natural way to include light quark masses in the hadron mass spectrum is to leave the LF potential unchanged  as a first approximation and add the additional term of the invariant mass $ \Delta m^2 = \sum_{i=1}^n \frac{m_i^2}{x_i}$ to the LF kinetic energy. The resulting LF wave function is then  modified by  the factor $e^{-\frac{1}{2\la} {\Delta m^2}}$, thus providing a relativistically invariant form  for the hadronic wave functions.  The effect of the nonzero quark masses for the squared hadron masses is then given by the expectation value of $ \Delta m^2$ evaluated using the modified wave functions. This prescription leads to the quadratic mass correction
\beq \label{DeM}
\De M^2[m_1,\cdots,m_n]=\frac{\lambda^2}{F}\,\frac{\ud F}{\ud\lambda},
\enq
with
$F[\lambda]=\int_0^1\cdots \int \ud x_1  \cdots \,\ud x_n
\,  e^{-\frac{1}{\la}\left(\sum_{i=1}^n
\frac{m_i^2}{x_i}\right)}\de(\sum_{i=1}^n x_i-1)$.

The resulting expressions for the squared masses of all light mesons and baryons are:
\beqa \label{mesfin}  
\mbox{Mesons} && M^2 = 4 \la (n+L)+ 2 \la \, s + \De M^2[m_1,m_2] ,\\
\label{barfin}\mbox{Baryons} && M^2=4 \la(n+L+1) + 2 \la \, s + \De M^2[m_1,m_2,m_3] ,\\
\label{tetrafin}   \mbox{Tetraquarks} && M^2=4 \la(n+L+1) + 2 \la \, s + \De M^2[m_1,m_2,m_3,m_4] ,
\enqa 
where the different values of the mass corrections within the supermultiplet break supersymmetry explicitly. For the tetraquark the mass formula is the same as for the baryon except for the quark mass correction  $\De M^2[m_1,m_2,m_3,m_4]$ given by Eq. \req{DeM}.

The pion mass of $\sim 0.140$ GeV is obtained if the non-strange light-quark mass is $m=0.045$ GeV~\cite{Brodsky:2014yha}.   In the case of the  $K$-meson,  the resulting value for the strange quark mass is $m_s=0.357$ GeV~\cite{Brodsky:2014yha}.    The trajectories of $K$, $K^*$ and $\phi$-mesons can then be readily calculated. (The predictions  are compared with experiment in Ref.~\cite{Brodsky:2014yha}. )  In Eq.~\req{DeM} the values of $x_i$ for the quarks are assumed to be uncorrelated. If one instead assumes maximal correlations in the cluster, i.e. $x_2=x_3$, this affects the final result by less than 1~\% for light quarks and less than 2~\% for the $\Omega^-$  which has three strange quarks. Therefore,  the  previously obtained agreement with the data~\cite{deTeramond:2014asa}  for the baryon spectra is hardly affected.   

One can fit the value of the fundamental mass parameter $\sqrt{\la}$ for each meson and baryon Regge  trajectory separately using Eqs.~\req{mesfin} and~\req{barfin} . The results are displayed in Fig.~\ref{slope}.   The best fit gives 
$\sqrt \lambda = 0.52$ GeV as the characteristic mass scale of QCD.

\begin{figure}[ht]
\setlength{\unitlength}{1mm}
\begin{center}
\includegraphics[width=15.8cm]{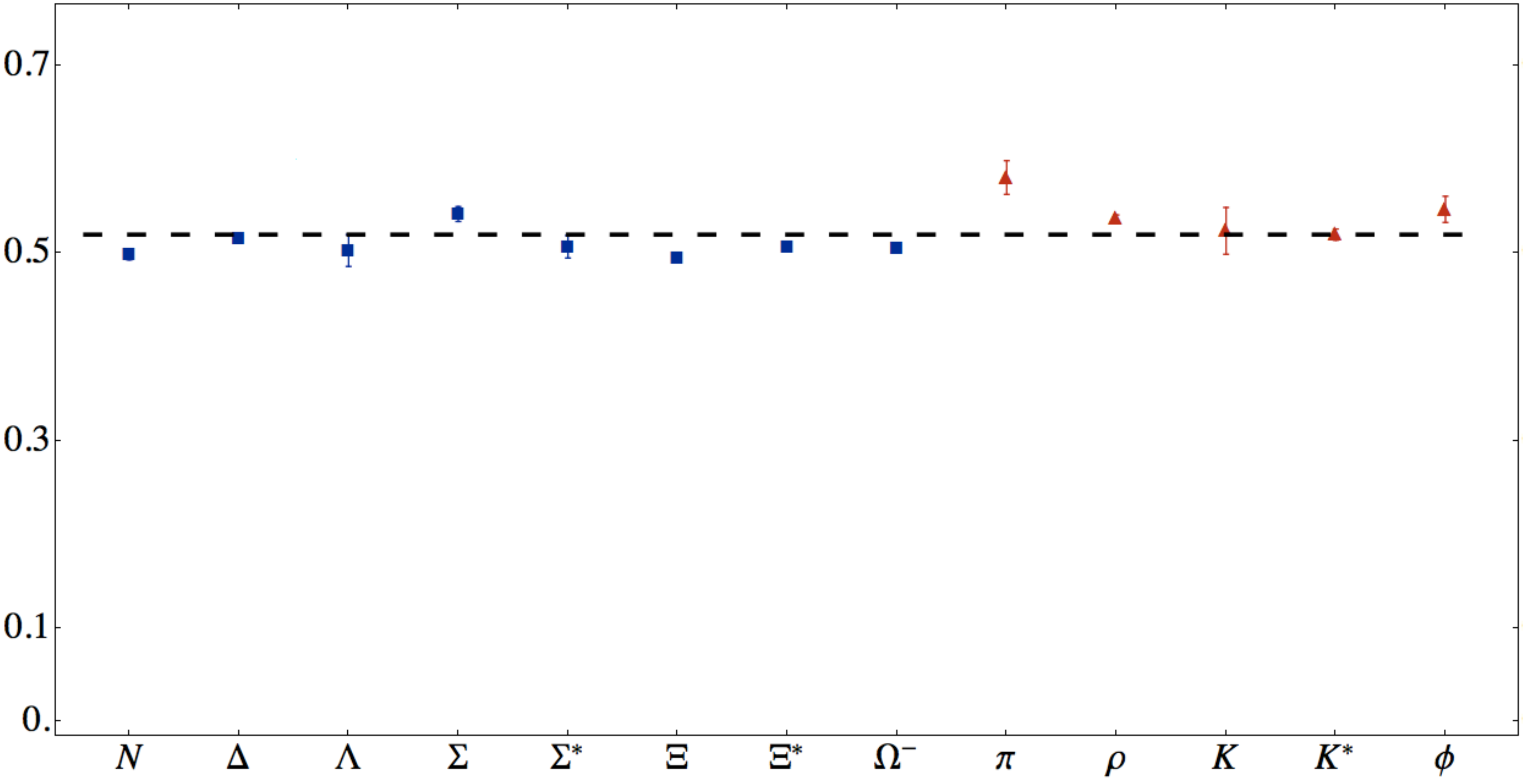}
\end{center}
\caption{\label{slope} \small Best fit for the value of the hadronic scale $\sqrt{\la}$ for the different Regge trajectories for baryons  (squares) and mesons (triangles) including all radial and orbital excitations using Eqs.~\req{mesfin} and~\req{barfin}.  The dotted line is the average value $\sqrt{\la}= 0.523$ GeV; it has the standard deviation  $\si =0.024$ GeV. For the  baryon sample alone the values are $0.509 \pm  0.015$ GeV and for the mesons $0.524 \pm 0.023$ GeV.}
\end{figure}

\section{Conclusions}

Inspired by the correspondence of classical gravitational theory  in 5-dimensional  AdS space with superconformal quantum field theory in physical 4-dimensional space-time, as originally proposed by Maldacena,  we have arrived at a novel holographic application  of supersymmetric quantum mechanics to light-front quantized Hamiltonian theory in physical space-time. The resulting superconformal algebra, which is the basis of our semiclassical theory, not only determines the breaking of the maximal symmetry of the dual gravitational theory, but it also provides the form of the frame-independent  color-confining light-front potential in the semiclassical theory.

Remarkably, supersymmetric quantum mechanics together with light-front holography account for principal features of hadron physics, such as the approximatively linear Regge trajectories (including the daughter trajectories) with nearly equal slopes for all mesons and baryons in both $L$  and $n$.  One finds remarkable supersymmetric aspects of hadron physics, connecting the masses of mesons and their superpartner baryons which are related by $L_M=L_B+1$.  The predicted spectroscopy for the meson and baryon superpartners agree with the data up to  an average absolute deviation of about 10 \%~\footnote{See also  Figs. 5.3 and 5.4 in Ref.~\cite{Brodsky:2014yha}, Figs. 2 and 3 in Ref.~\cite{Dosch:2015nwa},  and Fig. 1  in Ref.~\cite{Dosch:2015bca}.}. The agreement with experiment is generally better for the trajectories with total (cluster) spin $s=1$, such as the $\rho-\Delta$ superpartner trajectory  than for the trajectories with $s=0$, the $\pi-N$ trajectories. Features expected from spontaneous chiral symmetry  breaking are obtained, such as the masslessness of the pion in the massless quark limit. The structure of the superconformal algebra also implies that the pion has no supersymmetric partner. The meson-baryon supersymmetry survives, even if the heavy quark masses strongly break the conformal symmetry~\cite{Dosch:2015bca}.   In a subsequent publication we shall discuss the full supersymmetric quadruplets of heavy hadrons and especially exploit the fact that the boson and fermion potentials for heavy quarks can be derived from an {\it a priori} unknown but common supersymmetric potential. The diquark structure of heavy tetraquarks can be deduced from Table \ref{tetratable} by replacing a strange quark by a heavy one.

The structure of the hadronic mass generation obtained from the supersymmetric Hamiltonian $G_S$,  Eq. \req{GS},  provides  a frame-independent decomposition of the quadratic masses for all four members of the supersymmetric multiplet. In the massless quark limit:
\beq 
M^2_H/ \lambda =\overbrace{\underbrace{(2n + L_H +1)}_{\it kinetic} +\underbrace{(2n + L_H +1)}_{\it potential }}^{\small  \begin{array}{c} \mbox{\it  contribution 
from 2-dim}\\  \mbox{\it light-front harmonic oscillator}\end{array}} \hspace{6mm}
+   \hspace{-6mm} \overbrace{2(L_H+ s)  +2 \chi}^{\small \begin{array}{c} \mbox{\it contribution 
from AdS and }\\  \mbox{\it superconformal algebra}\end{array}} .
\enq
Here $n$ is the radial excitation number and  $L_H$ the LF angular momentum of the hadron wave function; $s$ is the total spin of the meson and  the cluster respectively,   $\chi = -1$ for the meson and for  the negative-chirality component of the baryon (the upper components in the susy-doublet) and $\chi=+1$ for the positive-chirality component  of baryon and for the tetraquark (the lower components).  
The contributions  to the hadron masses squared  from the light-front  potential $\lambda^2 \zeta^2$ and the light-front kinetic energy in the LF Hamiltonian, are identical because of the virial theorem.

We emphasize that the supersymmetric features of hadron physics derived here from superconformal quantum mechanics refers to the symmetry properties of  the bound-state wave functions of hadrons and not to quantum fields; there is therefore no need to introduce new supersymmetric  fields or particles such as squarks or gluinos.

We have argued that  tetraquarks --which are degenerate with the baryons with the same (leading) orbital angular momentum-- are required to complete the supermultiplets predicted by the superconformal algebra. The tetraquarks are the bound states of the same confined color-triplet diquarks and anti-diquarks which account for baryon spectroscopy.

The light-front cluster decomposition~\cite{Brodsky:1983vf,Brodsky:1985gs} for a bound state of $N$ constituents  --as an ``active" constituent interacting with the remaining cluster of $N-1$ constituents-- also has implications for the holographic description of form factors.  As a result, the form factor is written as the product of a two-body form factor multiplied by the form factor of the $N-1$ cluster evaluated at its characteristic scale. The form factor of the $N-1$ cluster is then expressed recursively in terms of the form factor of the $N-2$ cluster, and so forth, until the overall form factor is expressed as the $N-1$ product of two-body form factors evaluated at different characteristic scales. This cluster decomposition is in complete agreement with the QCD twist assignment which leads to counting-rule scaling laws~\cite{Brodsky:1973kr, Polchinski:2001tt}. This solves a previous problem with the twist assignment  for the nucleon~\cite{Brodsky:2014yha}: The ground state solution to the Hamiltonian~\req{Gsd} for the nucleon corresponds to twist 2: the nucleon is effectively described by the wave function of a quark-diquark cluster.  At short distances, however, all of the constituents in the proton are resolved, and therefore the falloff of the form factor at high $Q^2$ is governed by the total number of constituents; {\it i.e.}, it is twist 3~\footnote{A brief discussion of the LF cluster decomposition of form factors was given in Ref.~\cite{deTeramond:2016pov} and a will be discussed in more detail elsewhere.}. Also the twist assignment for the $\Delta$ (and total quark spin-$\frac{3}{2}$ nucleons) deviates from the assignment introduced in our previous papers~\cite{deTeramond:2013it, Brodsky:2014yha, deTeramond:2014asa, Dosch:2015nwa};  the approach chosen here, dictated by supersymmetry, does not require the introduction of half-integer twist.

The emerging confinement mass scale $\sqrt \la$ serves as the fundamental mass scale of QCD; it is directly related to physical observables such as  hadron masses and radii;  in addition, as discussed in Ref.~\cite{Deur:2014qfa},  it can be related to the scheme-dependent perturbative QCD scales, such as the QCD renormalization parameter $\La_s$.

\section*{Acknowledgments}

We thank Alexandre Deur for valuable discussions.

\end{document}